\documentclass{iaus}
\usepackage{graphicx}

\title[Infrared characteristics of sources associated with OH, H$_{2}$O,
SiO and CH$_{3}$OH masers] 
{Infrared characteristics of sources associated with OH, H$_{2}$O,
SiO and CH$_{3}$OH masers}
\author[Jarken. Esimbek, Jian Jun. Zhou, Gang. Wu, \& Xin Di. Tang]   
{Jarken. Esimbek$^1$$^{,}$$^3$,Jian Jun. Zhou$^1$$^{,}$$^3$, Gang.
Wu$^1$$^{,}$$^3$
\\ \and Xin Di. Tang$^1$$^{,}$$^2$}

\affiliation{$^1$Xinjiang Astronomical Observatory, Chinese Academy of Sciences, Urumqi 830011, PR China
 \\ email: {\tt jarken@xao.ac.cn}\\[\affilskip]
$^2$Graduate University of the Chinese Academy of Sciences,
Beijing 100080, PR China\\
$^3$Key Laboratory of Radio Astronomy, Chinese Academy of
Sciences, Urumqi 830011, PR China}

\pubyear{2012}
\volume{xxx}  
\pagerange{119--126}
\jname{Title of your IAU Symposium}
\editors{A.C. Editor, B.D. Editor \& C.E. Editor, eds.}
\begin{document}

\maketitle

\begin{abstract}
We collect all published OH, H$_{2}$O, SiO and CH$_{3}$OH masers in literature. The associated infrared sources of these four masers were identified with MSX PSC catalogues. We look for common infrared properties among the sources associated with four masers and make a statistical study. The MSX sources associated with stellar OH, stellar H$_{2}$O and SiO masers concentrated in a small regions and the MSX sources associated with interstellar OH, interstellar H$_{2}$O and CH$_{3}$OH masers also concentrated in a small regions in an [A]-[D].vs.[A][-[E] diagram. These results give us new criterion to search for coexisting stellar maser samples for OH, H$_{2}$O and SiO masers and interstellar maser samples for OH, H$_{2}$O and CH$_{3}$OH masers.

\end{abstract}

\firstsection 
\section{The sample}

We collect all published 1602 43 GHz SiO maser sources (Deguchi et al. 2004a, 2004b, 2007, 2010; Fujii et al. 2006; Jiang et al.
2002; Nakashima et al. 2003a, 2003b), 1712 6.7 GHz CH$_{3}$OH maser sources (Pestalozzi et al. 2005; Green et al. 2010; Xu et
al. 2003, 2010 ), 1417 22.235 GHz H$_{2}$O maser sources (Esimbek et al. 2005) and 3249 1612 MHz OH maser sources (Mu et al. 2010).

The MSX mission surveyed the entire Galactic belt within $¦òb¦ò\leq 4.5^{\circ}$ in five infrared bands B,A,C,D and E at 4,8,12,15 and 21 $\mu$m (Price et al. 2001). There are 1155 of the 1602 SiO masers, and 885 of the 1712 CH$_{3}$OH masers associated with MSX PSC sources within 1$^{\prime}$. Of the 743 H$_{2}$O masers associated with MSX PSC sources within 1$^{\prime}$, 300 are interstellar, and 155 are stellar masers. Of the 1869 OH masers associated with MSX PSC sources within 1$^{\prime}$, 63 are interstellar, and 1657 are stellar masers, others are unknown type.

\section{Statistical results }

OH, H$_{2}$O, SiO and CH$_{3}$OH are the most strong and widespread astrophysical masers. OH and H$_{2}$O masers could occur in star forming regions and the envelopes of evolved stars. Most of the SiO masers belong to circumstellar maser and CH$_{3}$OH masers are interstellar masers (Elitzur 1992). Fig1. Presents color indexes [A]-[D] vs. [A]-[E] associated with OH, H$_{2}$O, SiO and CH$_{3}$OH masers, respectively. Here, [A]-[D] and [A]-[E] denote log(F$_{D}$/F$_{A}$) and log (F$_{E}$/F$_{A}$). These mid-IR sources associated with these four masers are concentrated in small regions, while mid-IR sources associated with stellar OH masers are distributed in relatively larger region. Fig 2a. Presents color indexes [A]-[D] vs. [A]-[E] associated with stellar OH, stellar H$_{2}$O and SiO masers, there 66\%, 77\% and 95\% mid-IR sources associated with three masers are located in a area (-0.2,-0.2) ,(0.1,0.4),(0.4,0.4) and
(-0.2,-0.4). Fig 2b. Presents color indexes [A]-[D] vs. [A]-[E] associated with interstellar OH, interstellar H$_{2}$O and CH$_{3}$OH masers, there 81\%, 75\% and 92\% mid-IR sources associated with three masers are located in a area (-0.2,0.4) ,(0.5,1.5), (1,1.5) and (0.4,0.4).

\begin{figure}[h]
\begin{center}
 \includegraphics[width=3.4in]{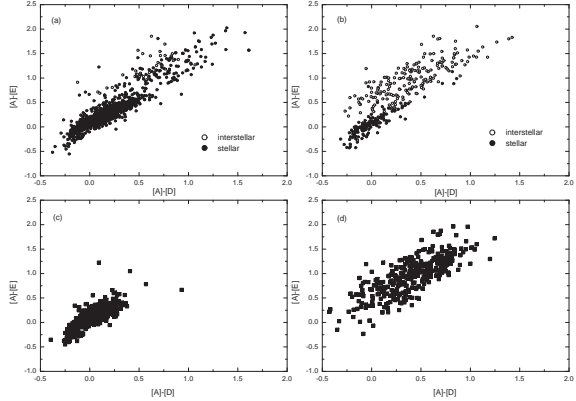}
 \caption{MSX color-color diagram of [A]-[D] vs. [A]-[E] associated with (a) OH (b) H$_{2}$O (c) SiO and (d) CH$_{3}$OH masers.}
   \label{fig1}
\end{center}
\end{figure}

\begin{figure}[h]
\begin{center}
 \includegraphics[width=3.2in]{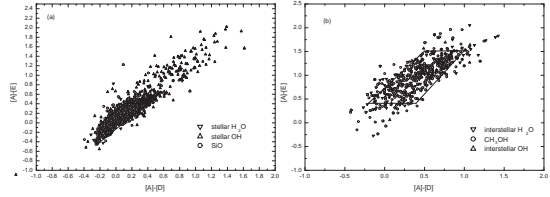}
 \caption{(a) MSX color-color diagram of [A]-[D] vs. [A]-[E] associated with stellar OH, stellar H$_{2}$O and SiO masers.
(b) MSX color-color diagram of [A]-[D] vs. [A]-[E] associated with
interstellar OH, interstellar H$_{2}$O and CH$_{3}$OH masers.}
   \label{fig2}
\end{center}
\end{figure}

\section{Summary}

We collect all published OH maser, H$_{2}$O maser, SiO maser and CH$_{3}$OH maser in literature and selected MSX PSC sources within
1$^{\prime}$ as the exciting sources. The mid-IR sources associated with stellar OH, stellar H$_{2}$O and SiO masers are concentrated in a area (-0.2,-0.2), (0.1,0.4), (0.4,0.4) and (-0.2,-0.4) and the mid-IR sources associated with interstellar OH, interstellar H$_{2}$O and CH$_{3}$OH masers are concentrated in a area (-0.2,0.4), (0.5,1.5), (1,1.5) and (0.4,0.4) in an [A]-[D] vs. [A]-[E](see Fig2.). This result provide us new criterion to search for coexisting stellar OH, stellar H$_{2}$O and SiO masers and coexisting interstellar OH, interstellar H$_{2}$O and CH$_{3}$OH masers.

Acknowledgements.
This work was funded by the National Natural Science foundation of China under grant 10778703 and 10873025, and partly supported by China Ministry of Science and Technology under State Key Development Program for Basic Research (2012CB821800).

\vspace{1mm}
 \scriptsize{

\end{document}